\documentclass[preprint]{ptephy}

\preprintnumber{KYUSHU-HET-163}

\usepackage{amssymb}
\usepackage{amsthm}
\usepackage{amsmath}
\usepackage{booktabs}
\usepackage{bbm}
\usepackage{mathtools}
\DeclareMathOperator{\tr}{tr}

\newcommand{\Slash}[1]{{\ooalign{\hfil/\hfil\crcr$#1$}}}
\numberwithin{equation}{section}

\usepackage{url}

\begin{document}

\title{Small flow-time representation of fermion bilinear operators}

\author{%
\name{\fname{Kenji} \surname{Hieda}}{1}
and
\name{\fname{Hiroshi} \surname{Suzuki}}{2,\ast}
}

\address{%
\affil{1,2}{
Department of Physics, Kyushu University
744 Motooka, Nishi-ku, Fukuoka, 819-0395, Japan}
\email{hsuzuki@phys.kyushu-u.ac.jp}
}

\begin{abstract}
Fermion bilinear operators of mass dimension~$3$, such as the axial-vector
and vector currents, the pseudo-scalar and scalar densities, whose
normalizations are fixed by Ward--Takahashi (WT) relations, are related to
small flow-time behavior of composite operators of fermion fields evolved by
L\"uscher's flow equation. The representations can be useful in lattice
numerical simulations, as recently demonstrated by the WHOT QCD collaboration
for the chiral condensation of the $N_f=2+1$ quantum chromodynamics (QCD) at
finite temperature.
\end{abstract}
\subjectindex{B01, B31, B32, B38}
\maketitle

\section{Introduction}
\label{sec:1}
It is a remarkable fact~\cite{Luscher:2011bx,Luscher:2013cpa} that, under the
conventional parameter renormalization in gauge theory, a simple one-parameter
evolution of bare gauge and bare fermion fields by flow equations
in~Refs.~\cite{Luscher:2009eq,Luscher:2010iy,Luscher:2013cpa} renders composite
operators of these fields renormalized.\footnote{Up to multiplicative
renormalization factors determined by the number of fermion fields contained in
a composite operator. The use of ``ringed fermion variables''
in~Eqs.~\eqref{eq:(2.11)} and~\eqref{eq:(2.12)} avoids this renormalization.}
See also Ref.~\cite{Narayanan:2006rf} for an earlier study. Because of this
renormalization property, the flow can be employed in lattice gauge theory for
instance to set a mass scale, to define a non-perturbative running coupling,
and to define the topological charge.
Refs.~\cite{Luscher:2013vga,Ramos:2015dla} are reviews and
Ref.~\cite{Hieda:2016xpq} provides a pedestrian proof of the above
renormalization property and an extensive list of recent related works.

Although any composite operator of fields evolved by the flow (which we term
``flowed fields'' in what follows) is always a renormalized one, how to relate
it to desired renormalized quantities in the original gauge theory is a
different issue. One possible rather versatile method to do this is the small
flow-time expansion~\cite{Luscher:2011bx}. The flow is parametrized by the
flow-time~$t>0$ and the small flow-time expansion expresses for~$t\sim0$ a
composite operator of flowed fields by an asymptotic series of local operators.
Since the flow-time~$t$ possesses the mass dimension~$-2$, the expansion is a
series of local operators of increasing mass dimensions. On the basis of this
small flow-time expansion, representations of renormalized Noether currents,
such as the energy--momentum tensor~\cite{Suzuki:2013gza,Makino:2014taa} and
the flavor non-singlet axial vector current~\cite{Endo:2015iea}, in terms of
flowed fields have been constructed. Since composite operators of flowed fields
are renormalized ones and independent of the regularization, those
representations are thought to be useful in lattice gauge theory. The validity
of the representation of the energy--momentum tensor has been numerically
examined in~Refs.~\cite{Asakawa:2013laa,Kitazawa:2014uxa,Itou:2015gxx,%
Kitazawa:2015ewp,Kitazawa:2016,Taniguchi:2016} with promising results.

In this paper, as further extension of the above idea, we construct a small
flow-time representation of various fermion bilinear operators of mass
dimension~$3$. These include the axial-vector and vector currents, the
pseudo-scalar and scalar densities, both flavor non-singlet and flavor singlet
ones. Our basic principle is that the normalizations of the operators are fixed
by Ward--Takahashi (WT) relations associated with the relevant symmetry.
Although it is not a \textit{priori\/} clear whether those small flow-time
representations are practically useful, a preliminary numerical study of the
chiral condensation in the $N_f=2+1$ QCD at finite
temperature~\cite{Taniguchi:2016} on the basis of the small flow-time
representation suggests that our representations are practically useful; this
observation encouraged us to publish the present work.

In this paper, we consider the vector-like gauge theory with $N_f$~flavor
fermions. The classical action is given by\footnote{Here is our notation:
Without noting otherwise, repeated indices are understood to be summed over.
The generators~$T^a$ of the gauge group~$G$ are anti-hermitian and the
structure constants are defined by~$[T^a,T^b]=f^{abc}T^c$. Quadratic Casimirs
are defined by~$f^{acd}f^{bcd}=C_2(G)\delta^{ab}$ and, for a gauge
representation~$R$,
$\tr_R(T^aT^b)=-T(R)\delta^{ab}$ and~$T^aT^a=-C_2(R)1$. We also denote
$\tr_R(1)=\dim(R)$. For the fundamental $N$~representation of~$SU(N)$ for which
$\dim(N)=N$, our choice is
\begin{equation}
   C_2(SU(N))=N,\qquad T(N)=\frac{1}{2},\qquad
   C_2(N)=\frac{N^2-1}{2N}.
\label{eq:(1.1)}
\end{equation}
Our $\gamma$ matrices are all Hermitian and for the trace over the spinor index
we set $\tr(1)=4$ for any spacetime dimension~$D$. The chiral matrix is defined
by~$\gamma_5=\gamma_0\gamma_1\gamma_2\gamma_3$ for any~$D$ and thus
\begin{equation}
   \tr(\gamma_5\gamma_\mu\gamma_\nu\gamma_\rho\gamma_\sigma)
   =\begin{cases}
   4\epsilon_{\mu\nu\rho\sigma},&\mu,\nu,\rho,\sigma\in\{0,1,2,3\},\\
   0,&\text{otherwise},
   \end{cases}
\label{eq:(1.2)}
\end{equation}
where the totally anti-symmetric tensor is normalized as~$\epsilon_{0123}=1$.}
\begin{equation}
   S=\frac{1}{4g_0^2}\int d^Dx\,F_{\mu\nu}^a(x)F_{\mu\nu}^a(x)
   +\int d^Dx\,\Bar{\psi}(x)(\Slash{D}+M_0)\psi(x),
\label{eq:(1.3)}
\end{equation}
where $g_0$ and~$M_0$ are the bare gauge coupling and the bare mass matrix,
respectively; the flavor index is almost always suppressed and we assume that
the mass matrix~$M_0$ is flavor-diagonal. The field strength is defined by
\begin{equation}
   F_{\mu\nu}(x)
   =\partial_\mu A_\nu(x)-\partial_\nu A_\mu(x)+[A_\mu(x),A_\nu(x)],
\label{eq:(1.4)}
\end{equation}
from the gauge potential~$A_\mu(x)=A_\mu^a(x)T^a$
and~$F_{\mu\nu}(x)=F_{\mu\nu}^a(x)T^a$. The covariant derivative on the fermion
is given by
\begin{equation}
   D_\mu=\partial_\mu+A_\mu.
\label{eq:(1.5)}
\end{equation}

Our flow equations are identical to those
of~Refs.~\cite{Luscher:2009eq,Luscher:2010iy,Luscher:2013cpa}. The flow of the
gauge field along the flow-time~$t$ is defined by
\begin{equation}
   \partial_tB_\mu(t,x)=D_\nu G_{\nu\mu}(t,x),\qquad
   B_\mu(t=0,x)=A_\mu(x),
\label{eq:(1.6)}
\end{equation}
where
\begin{equation}
   D_\mu\equiv\partial_\mu+[B_\mu,\cdot],\qquad
   G_{\mu\nu}(t,x)
   \equiv\partial_\mu B_\nu(t,x)-\partial_\nu B_\mu(t,x)
   +[B_\mu(t,x),B_\nu(t,x)],
\label{eq:(1.7)}
\end{equation}
and the flow for the fermion fields is defined by
\begin{align}
   \partial_t\chi(t,x)&=D^2\chi(t,x),&
   \chi(t=0,x)&=\psi(x),
\label{eq:(1.8)}
\\
   \partial_t\Bar{\chi}(t,x)
   &=\Bar{\chi}(t,x)
   \overleftarrow{D}^2,&
   \Bar{\chi}(t=0,x)&=\Bar{\psi}(x),
\label{eq:(1.9)}
\end{align}
where the covariant derivatives on the fermion fields are defined by
\begin{equation}
   D_\mu=\partial_\mu+B_\mu,\qquad
   \overleftarrow{D}_\mu\equiv\overleftarrow{\partial}_\mu-B_\mu.
\label{eq:(1.10)}
\end{equation}

\section{Small flow-time representation of fermion bilinear operators}
\label{sec:2}
\subsection{One-loop coefficients in the small flow-time expansion}
\label{sec:2.1}
Because of the gauge and flavor symmetries, the small flow-time
expansion~\cite{Luscher:2011bx} of a dimension~$3$ fermion bilinear operator
assumes the form
\begin{equation}
   \Bar{\chi}(t,x)\mathcal{M}\chi(t,x)
   \stackrel{t\to0}{\sim}
   \zeta_0(t)\mathbbm{1}
   +\zeta_1(t)\Bar{\psi}(x)\mathcal{M}\psi(x)+O(t),
\label{eq:(2.1)}
\end{equation}
where $\mathcal{M}$ is a certain constant matrix in the spinor and flavor
spaces and $\mathbbm{1}$ denotes the identity operator. The coefficients
in~Eq.~\eqref{eq:(2.1)} are expanded in the loop expansion as
\begin{equation}
   \zeta_0(t)=\sum_{\ell=1}^\infty\zeta_0^{(\ell)}(t),\qquad
   \zeta_1(t)=1+\sum_{\ell=1}^\infty\zeta_1^{(\ell)}(t).
\label{eq:(2.2)}
\end{equation}

\begin{figure}[ht]
\begin{minipage}{0.3\columnwidth}
\begin{center}
\includegraphics[width=0.6\columnwidth]{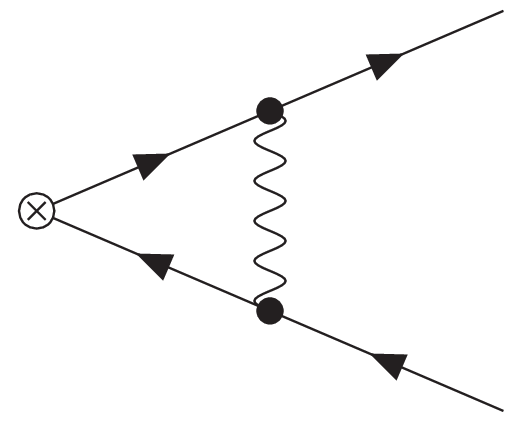}
\end{center}
\end{minipage}
\begin{minipage}{0.3\columnwidth}
\begin{center}
\includegraphics[width=0.6\columnwidth]{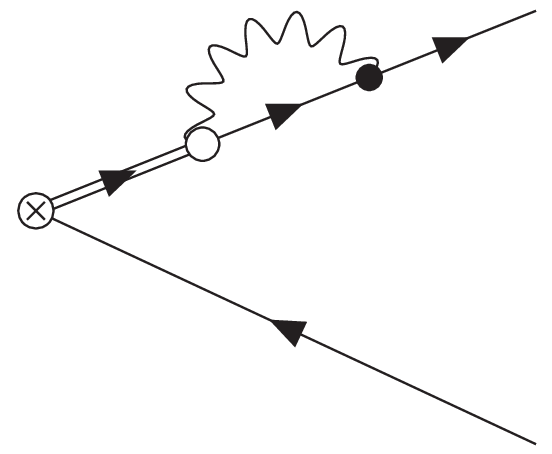}
\end{center}
\end{minipage}
\begin{minipage}{0.3\columnwidth}
\begin{center}
\includegraphics[width=0.6\columnwidth]{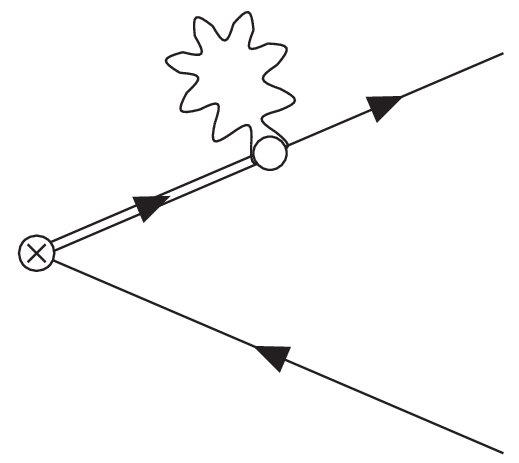}
\end{center}
\end{minipage}
\caption{The flow Feynman diagrams which contribute to Eq.~\eqref{eq:(2.4)}.
The blob denotes the composite operator in the left-hand side
of~Eq.~\eqref{eq:(2.1)}. The convention here is identical to that
in~Refs.~\cite{Makino:2014taa,Endo:2015iea}: The single straight line and the
doubled straight line denote the fermion propagator and the fermion heat
kernel, respectively; the wavy line denotes the gauge propagator. The filled
circle denotes the vertex in the original gauge theory, while the open circle
denotes the vertex originating from non-linear terms in the fermion flow
equations.}
\label{fig:1}
\end{figure}

The computation of the one-loop coefficients $\zeta_0^{(1)}$ and~$\zeta_1^{(1)}$
is straightforward; see~Refs.~\cite{Makino:2014taa,Endo:2015iea,Suzuki:2015bqa}
for actual one-loop calculations. Assuming the dimensional regularization
with~$D=4-2\epsilon$, we find
\begin{equation}
   \zeta_0^{(1)}(t)
   =-\frac{1}{(4\pi)^2}\dim(R)\tr
   \left(\mathcal{M}M
   \left\{
   \frac{1}{2t}
   +M^2\left[\gamma+\ln(2M^2 t)\right]
   +O(t)
   \right\}
   \right),
\label{eq:(2.3)}
\end{equation}
where $M=M_0[1+O(g^2)]$ is the renormalized mass matrix, $\gamma$ denotes the
Euler constant, and
\begin{equation}
   \zeta_1^{(1)}(t)
   =\frac{g^2}{(4\pi)^2}C_2(R)
   \begin{Bmatrix*}[l]
   (-6)\left[\dfrac{1}{\epsilon}+\ln(8\pi\mu^2t)\right]-2
   & \text{when $\mathcal{M}\propto 1$}\\
   (-3)\left[\dfrac{1}{\epsilon}+\ln(8\pi\mu^2t)\right]+\dfrac{1}{2}
   & \text{when $\mathcal{M}\propto \gamma_\mu$}\\
   (-2)\left[\dfrac{1}{\epsilon}+\ln(8\pi\mu^2t)\right]
   & \text{when $\mathcal{M}\propto \gamma_{[\mu}\gamma_{\nu]}$}\\
   (-3)\left[\dfrac{1}{\epsilon}+\ln(8\pi\mu^2t)\right]-\dfrac{7}{2}
   & \text{when $\mathcal{M}\propto \gamma_\mu\gamma_5$}\\
   (-6)\left[\dfrac{1}{\epsilon}+\ln(8\pi\mu^2t)\right]-10
   & \text{when $\mathcal{M}\propto \gamma_5$}\\
   \end{Bmatrix*}+O(t),
\label{eq:(2.4)}
\end{equation}
where we have used the gauge coupling renormalization
$g_0^2=\mu^{2\epsilon}g^2[1+O(g^2)]$. This is obtained by evaluating the flow
Feynman diagrams depicted in~Fig.~\ref{fig:1}.

\subsection{Flavor non-singlet axial-vector current and the pseudo-scalar density}
\label{sec:2.2}
Our first example is the flavor non-singlet axial-vector
current~$j_{5\mu}^A(x)$ and the flavor non-singlet pseudo-scalar density. These
have been already considered in~Ref.~\cite{Endo:2015iea} but we recapitulate
the basic reasoning and the results here for completeness and for later use.

\begin{figure}[ht]
\centering
\includegraphics[width=0.9\columnwidth]{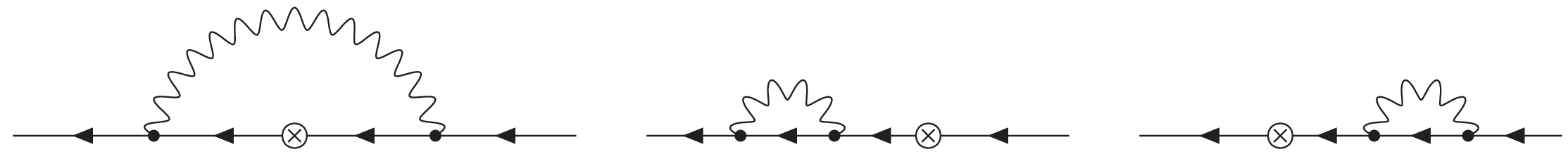}
\caption{The one-loop diagrams which contribute to the chiral WT identity with
two external fermion lines. The blob stands for a composite operator.}
\label{fig:2}
\end{figure}

The correctly normalized flavor non-singlet axial-vector
current~$j_{5\mu}^A(x)$ and the flavor non-singlet renormalized pseudo-scalar
density are characterized by the following chiral WT relation (the partially
conserved axial-vector current (PCAC) relation):\footnote{In writing this
expression, we assume that the gauge fixing is made.}
\begin{align}
   &\left\langle
   \left[
   \partial_\mu j_{5\mu}^A(x)
   -\left\{\Bar{\psi}\gamma_5\{t^A,M\}\psi\right\}_R(x)\right]
   \prod_i\psi(y_i)\prod_j\Bar{\psi}(z_j)\prod_kA_{\mu_k}^{a_k}(w_k)
   \right\rangle
\notag\\
   &=-\sum_l\delta(x-y_l)
   \left.\left\langle
   \prod_i\psi(y_i)\prod_j\Bar{\psi}(z_j)\prod_kA_{\mu_k}^{a_k}(w_k)
   \right\rangle\right|_{\psi(y_l)\to\gamma_5t^A\psi(y_l)}
\notag\\
   &\qquad\qquad{}
   -\sum_l\delta(x-z_l)
   \left.\left\langle
   \prod_i\psi(y_i)\prod_j\Bar{\psi}(z_j)\prod_kA_{\mu_k}^{a_k}(w_k)
   \right\rangle\right|_{\Bar{\psi}(z_l)\to\Bar{\psi}(z_l)\gamma_5t^A},
\label{eq:(2.5)}
\end{align}
where $t^A$ denotes the (anti-hermitian) generator of the flavor group. An
analysis of the one-loop diagrams in~Fig.~\ref{fig:2} shows that, under the
dimensional regularization, such operators are given by (see \S13.5
of~Ref.~\cite{Collins:1984xc})
\begin{align}
   j_{5\mu}^A(x)
   &=\left[1+\frac{g^2}{(4\pi)^2}C_2(R)(-4)+O(g^4)\right]
   \Bar{\psi}(x)\gamma_\mu\gamma_5t^A\psi(x),
\label{eq:(2.6)}
\\
   \left\{\Bar{\psi}\gamma_5\{t^A,M\}\psi\right\}_R(x)
   &=\left[1+\frac{g^2}{(4\pi)^2}C_2(R)(-8)+O(g^4)\right]
   \Bar{\psi}(x)\gamma_5\{t^A,M_0\}\psi(x).
\label{eq:(2.7)}
\end{align}
We thus express bare operators on the right-hand sides of these expressions by
composite operators of flowed fermion fields. This can be carried out by
inverting the small flow-time expansion~\eqref{eq:(2.1)}
with~Eq.~\eqref{eq:(2.2)} for $t$ small. Using Eq.~\eqref{eq:(2.4)}, we have
\begin{align}
   &j_{5\mu}^A(x)
   =\left\{1+\frac{g^2}{(4\pi)^2}C_2(R)
   3\left[\dfrac{1}{\epsilon}+\ln(8\pi\mu^2t)-\dfrac{1}{6}\right]
   +O(g^4)\right\}
   \Bar{\chi}(t,x)\gamma_\mu\gamma_5t^A\chi(t,x)+O(t),
\label{eq:(2.8)}
\\
   &\left\{\Bar{\psi}\gamma_5\{t^A,M\}\psi\right\}_R(x)
\notag\\
   &=\left\{1+\frac{g^2}{(4\pi)^2}C_2(R)
   6\left[\dfrac{1}{\epsilon}+\ln(8\pi\mu^2t)+\frac{1}{3}\right]
   +O(g^4)\right\}
   \Bar{\chi}(t,x)\gamma_5\{t^A,M_0\}\chi(t,x)+O(t).
\label{eq:(2.9)}
\end{align}

We further rewrite the resulting expressions in terms of the renormalized mass
in the minimal subtraction (MS) scheme,
\begin{equation}
   M_0=\left[1+\frac{g^2}{(4\pi)^2}C_2(R)(-3)\frac{1}{\epsilon}+O(g^4)
   \right]M,
\label{eq:(2.10)}
\end{equation}
and the ``ringed variables''~\cite{Makino:2014taa}
\begin{align}
   \mathring{\chi}_r(t,x)
   &=\sqrt{\frac{-2\dim(R)}
   {(4\pi)^2t^2
   \left\langle\Bar{\chi}_r(t,x)\overleftrightarrow{\Slash{D}}\chi_r(t,x)
   \right\rangle}}
   \,\chi_r(t,x),
\label{eq:(2.11)}\\
   \mathring{\Bar{\chi}}_r(t,x)
   &=\sqrt{\frac{-2\dim(R)}
   {(4\pi)^2t^2
   \left\langle\Bar{\chi}_r(t,x)\overleftrightarrow{\Slash{D}}\chi_r(t,x)
   \right\rangle}}
   \,\Bar{\chi}_r(t,x),
\label{eq:(2.12)}
\end{align}
where $\overleftrightarrow{D}_\mu\equiv D_\mu-\overleftarrow{D}_\mu$ and it is
understood that the sum over the flavor index~$r$ is not taken, by
using~\cite{Makino:2014taa}
\begin{align}
   &\frac{-2\dim(R)}
   {(4\pi)^2t^2\left\langle\Bar{\chi}_r(t,x)
   \overleftrightarrow{\Slash{D}}\chi_r(t,x)\right\rangle}
\notag\\
   &=\frac{1}{(8\pi t)^\epsilon}
   \left\{1+\frac{g^2}{(4\pi)^2}C_2(R)\left[3\frac{1}{\epsilon}
   +3\ln(8\pi\mu^2t)-\ln(432)\right]
   +O(t)+O(g^4)\right\}.
\label{eq:(2.13)}
\end{align}

In this way, we have
\begin{align}
   &j_{5\mu}^A(x)
   =\left\{1+\frac{g^2}{(4\pi)^2}C_2(R)
   \left[-\dfrac{1}{2}+\ln(432)\right]
   +O(g^4)\right\}
   \mathring{\Bar{\chi}}(t,x)\gamma_\mu\gamma_5t^A\mathring{\chi}(t,x)+O(t),
\label{eq:(2.14)}
\\
   &\left\{\Bar{\psi}\gamma_5\{t^A,M\}\psi\right\}_R(x)
\notag\\
   &=\left\{1+\frac{g^2}{(4\pi)^2}C_2(R)
   \left[3\ln(8\pi\mu^2t)+2+\ln(432)\right]
   +O(g^4)\right\}
   \mathring{\Bar{\chi}}(t,x)\gamma_5\{t^A,M\}\mathring{\chi}(t,x)+O(t).
\label{eq:(2.15)}
\end{align}
The ringed variables, $\mathring{\chi}(t,x)$ and~$\mathring{\Bar{\chi}}(t,x)$,
are free from the wave function renormalization of flowed fermion
fields~\cite{Luscher:2013cpa} and the above expressions~\eqref{eq:(2.14)}
and~\eqref{eq:(2.15)} are manifestly finite as they should be.

Finally, a renormalization group argument~\cite{Makino:2014taa,Endo:2015iea}
says that when renormalized parameters in these expressions are replaced by
running parameters, $g\to\Bar{g}(\mu)$ and~$M\to\Bar{M}(\mu)$, then the
expressions are independent of the renormalization scale~$\mu$. By setting
$\mu=1/\sqrt{8t}$, perturbation theory is justified in the limit~$t\to0$ owing
to the asymptotic freedom. Thus, we have
\begin{align}
   &j_{5\mu}^A(x)
   =\lim_{t\to0}\left\{1+\frac{\Bar{g}(1/\sqrt{8t})^2}{(4\pi)^2}C_2(R)
   \left[-\dfrac{1}{2}+\ln(432)\right]\right\}
   \mathring{\Bar{\chi}}(t,x)\gamma_\mu\gamma_5t^A\mathring{\chi}(t,x),
\label{eq:(2.16)}
\\
   &\left\{\Bar{\psi}\gamma_5\{t^A,M\}\psi\right\}_R(x)
\notag\\
   &=\lim_{t\to0}\left\{1+\frac{\Bar{g}(1/\sqrt{8t})^2}{(4\pi)^2}C_2(R)
   \left[3\ln\pi+2+\ln(432)\right]
   \right\}
   \mathring{\Bar{\chi}}(t,x)\gamma_5\{t^A,\Bar{M}(1/\sqrt{8t})\}\}
   \mathring{\chi}(t,x).
\label{eq:(2.17)}
\end{align}
These are expressions obtained in~Ref.~\cite{Endo:2015iea}. If one adopts the
modified MS  ($\overline{\text{MS}}$) scheme instead of the MS scheme, the
factor $\ln\pi$ in the last expression is replaced by~$\gamma-2\ln2$.

\subsection{Flavor non-singlet vector current}
\label{sec:2.3}
Our next example is the flavor non-singlet vector current. The flavor
non-singlet vector current~$j_\mu^A(x)$ is characterized by the WT relation,
\begin{align}
   &\left\langle
   \partial_\mu j_\mu^A(x)
   \prod_i\psi(y_i)\prod_j\Bar{\psi}(z_j)\prod_kA_{\mu_k}^{a_k}(w_k)
   \right\rangle
\notag\\
   &=-\sum_l\delta(x-y_l)
   \left.\left\langle
   \prod_i\psi(y_i)\prod_j\Bar{\psi}(z_j)\prod_kA_{\mu_k}^{a_k}(w_k)
   \right\rangle\right|_{\psi(y_l)\to t^A\psi(y_l)}
\notag\\
   &\qquad\qquad{}
   -\sum_l\delta(x-z_l)
   \left.\left\langle
   \prod_i\psi(y_i)\prod_j\Bar{\psi}(z_j)\prod_kA_{\mu_k}^{a_k}(w_k)
   \right\rangle\right|_{\Bar{\psi}(z_l)\to-\Bar{\psi}(z_l)t^A}.
\label{eq:(2.18)}
\end{align}
The dimensional regularization preserves the corresponding $SU(N_f)_V$ symmetry
and thus the correctly normalized current is simply given by
\begin{equation}
   j_\mu^A(x)=\Bar{\psi}(x)\gamma_\mu t^A\psi(x),
\label{eq:(2.19)}
\end{equation}
assuming the dimensional regularization.

Thus from~Eqs.~\eqref{eq:(2.1)}, \eqref{eq:(2.2)}, \eqref{eq:(2.4)},
\eqref{eq:(2.11)}, \eqref{eq:(2.12)}, and~\eqref{eq:(2.13)}, in conjunction of
the renormalization group argument, we have
\begin{equation}
   j_\mu^A(x)
   =\lim_{t\to0}\left\{1+\frac{\Bar{g}(1/\sqrt{8t})^2}{(4\pi)^2}C_2(R)
   \left[-\dfrac{1}{2}+\ln(432)\right]
   \right\}
   \mathring{\Bar{\chi}}(t,x)\gamma_\mu t^A\mathring{\chi}(t,x).
\label{eq:(2.20)}
\end{equation}

Now, there is another interesting method to obtain the
expression~\eqref{eq:(2.20)}. We may require that the flavor non-singlet
axial-vector current~$j_{5\mu}^A(x)$ and the flavor non-singlet vector
current~$j_\mu^A(x)$ form a current algebra. We can use this requirement as the
\emph{definition\/} of the vector current. That is, we would require that the
vector current of the form
\begin{equation}
   \left\{\Bar{\psi}\gamma_\mu[t^A,t^B]\psi\right\}_R(x),
\label{eq:(2.21)}
\end{equation}
is given by the infinitesimal chiral transformation
\begin{equation}
   \psi(x)\to(1+\alpha\gamma_5t^B)\psi(x),\qquad
   \Bar{\psi}(x)\to\Bar{\psi}(x)(1+\alpha\gamma_5t^B),
\label{eq:(2.22)}
\end{equation}
of the axial-vector current~$j_{5\mu}^A(x)$~\eqref{eq:(2.16)}.

The chiral transformation~\eqref{eq:(2.22)}, through the initial
conditions of the flow equations~\eqref{eq:(1.8)} and~\eqref{eq:(1.9)},
induces the chiral transformation on the flowed fermions fields,
\begin{equation}
   \chi(t,x)\to(1+\alpha\gamma_5t^B)\chi(t,x),\qquad
   \Bar{\chi}(t,x)\to\Bar{\chi}(t,x)(1+\alpha\gamma_5t^B).
\label{eq:(2.23)}
\end{equation}
We now claim that the composite operator of the flowed fields on the
right-hand side of~Eq.~\eqref{eq:(2.16)} transforms in a \emph{naive way\/}
under the chiral transformation~\eqref{eq:(2.23)}. That is, we claim that no
nontrivial renormalization is required under the transformation. That a
composite operator of the flowed fermion fields, such
as~$\Bar{\chi}(t,x)\gamma_\mu\gamma_5t^A\chi(t,x)$, transforms in a naive way
under the flavor non-singlet chiral transformation is explained in detail
in~Sect.~4.1 and Sect.~4.2 of~Ref.~\cite{Luscher:2013cpa}; see Eqs.~(4.14)
and~(4.15) there. One possible explanation is that the meaning of a composite
operator such as
$\mathring{\Bar{\chi}}(t,x)\gamma_\mu\gamma_5t^A\mathring{\chi}(t,x)$ is
independent of the regularization. Then one may adopt regularization (such as
the overlap fermion) that manifestly preserves the flavor non-singlet chiral
symmetry. Then the transformation law of such a composite operator can be
derived in a naive way.

Applying the chiral transformation~\eqref{eq:(2.22)} to both sides
of~Eq.~\eqref{eq:(2.16)} thus would yield
\begin{align}
   &\left\{\Bar{\psi}\gamma_\mu[t^A,t^B]\psi\right\}_R(x)
\notag\\
   &=\lim_{t\to0}\left\{1+
   \frac{\Bar{g}(1/\sqrt{8t})^2}{(4\pi)^2}C_2(R)
   \left[-\frac{1}{2}+\ln(432)\right]\right\}
   \mathring{\Bar{\chi}}(t,x)\gamma_\mu[t^A,t^B]\mathring{\chi}(t,x).
\label{eq:(2.24)}
\end{align}
By setting $[t^A,t^B]\to t^A$, this coincides with~Eq.~\eqref{eq:(2.20)} which
was obtained by a direct small flow-time expansion. This coincidence supports
the above argument on the basis of the naive transformation law of composite
operators of flowed fields.

\subsection{Flavor singlet vector current}
\label{sec:2.4}
For the flavor singlet vector current (i.e., the fermion number current)
\begin{equation}
   j_\mu(x)=\left\{\Bar{\psi}\gamma_\mu\psi\right\}_R(x),
\label{eq:(2.25)}
\end{equation}
the argument in the second half of the preceding subsection does not apply,
because this current is not related to the flavor non-singlet axial-vector
current (which fulfills the WT relation without anomaly) by the chiral
transformation. Still, since the expression
$j_\mu(x)=\Bar{\psi}(x)\gamma_\mu\psi(x)$ with the dimensional regularization
provides the correctly normalized current, we can directly use the small
flow-time expansion. Again from~Eqs.~\eqref{eq:(2.1)}, \eqref{eq:(2.2)},
\eqref{eq:(2.4)}, etc., we have
\begin{equation}
   j_\mu(x)
   =\lim_{t\to0}\left\{
   1+\frac{\Bar{g}(1/\sqrt{8t})^2}{(4\pi)^2}C_2(R)
   \left[-\frac{1}{2}+\ln(432)\right]
   \right\}
   \mathring{\Bar{\chi}}(t,x)\gamma_\mu\mathring{\chi}(t,x).
\label{eq:(2.26)}
\end{equation}

\subsection{Scalar density}
\label{sec:(2.5)}
As for the vector current~\eqref{eq:(2.24)}, we may obtain the small flow-time
representation of the scalar density of the form
\begin{equation}
   \left\{\Bar{\psi}\{\{t^A,M\},t^B\}\psi\right\}_R(x)
\label{eq:(2.27)}
\end{equation}
by requiring it to be obtained by the chiral transformation~\eqref{eq:(2.22)}
of the correctly normalized pseudo-scalar density~\eqref{eq:(2.17)}. Apart from
the chiral limit~$M\to0$, however, there exists a subtlety in the small
flow-time limit\footnote{We would like to thank Tetsuya Onogi and Hidenori
Fukaya for discussion regarding this subtlety.} and our argument should be
somewhat modified as follows.

We assume that the composite operator of the flowed fermion fields
in~Eq.~\eqref{eq:(2.15)} transforms in a naive way under the chiral
transformation. This assumption is the same as a previous subsection and, in
terms of the integrated form of the chiral WT relation~\eqref{eq:(2.5)}, this
is expressed as
\begin{align}
   &\int_{\partial\mathcal{R}}d^{D-1}y_\mu\,\left\langle
   j_{5\mu}^A(y)
   \mathring{\Bar{\chi}}(t,x)\gamma_5\{t^A,M\}\mathring{\chi}(t,x)
   \dotsb\right\rangle
\notag\\
   &\qquad{}
   -\int_{\mathcal{R}}d^Dy\,\left\langle
   \left\{\Bar{\psi}\gamma_5\{t^B,M\}\psi\right\}_R(y)
   \mathring{\Bar{\chi}}(t,x)\gamma_5\{t^A,M\}\mathring{\chi}(t,x)
   \dotsb\right\rangle
\notag\\
   &=-\left\langle
   \mathring{\Bar{\chi}}(t,x)\{\{t^A,M\},t^B\}\mathring{\chi}(t,x)
   \dotsb\right\rangle,
\label{eq:(2.28)}
\end{align}
where the bounded integration region~$\mathcal{R}$ contains the point~$x$.
$\partial\mathcal{R}$ is the boundary of the region~$\mathcal{R}$ and
$dy^\mu$ is the area element on~$\partial\mathcal{R}$. The abbreviated terms
($\dotsb$) stand for other operators contained in the correlation function. The
small flow-time expansion of the operator on the right-hand side yields,
from~Eqs.~\eqref{eq:(2.1)}, \eqref{eq:(2.2)}, and~\eqref{eq:(2.3)},
\begin{align}
   &\mathring{\Bar{\chi}}(t,x)\{\{t^A,M\},t^B\}\mathring{\chi}(t,x)
\notag\\
   &\stackrel{t\to0}{\sim}
   \left[-\frac{1}{(4\pi)^2}4\dim(R)\frac{1}{2t}\tr(\{\{t^A,M\},t^B\}M)
   +O(t^0)+O(g^2)\right]
   \mathbbm{1}
\notag\\
   &\qquad{}
   +[1+O(g^2)]\Bar{\psi}(x)\{\{t^A,M\},t^B\}\psi(x)+O(t).
\label{eq:(2.29)}
\end{align}
The first term on the right-hand side which is proportional to the identity
operator~$\mathbbm{1}$ behaves as~$\sim1/t$ for~$t\to0$.
In~Eq.~\eqref{eq:(2.28)}, although the small flow-time expansion of the
pseudo-scalar density
$\mathring{\Bar{\chi}}(t,x)\gamma_5\{t^A,M\}\mathring{\chi}(t,x)$ on the
left-hand side starts from an $O(t^0)$ term as Eq.~\eqref{eq:(2.9)} shows, the
equal point ($y=x$) product in the second line makes the right-hand side more
singular as~$t\to0$.

Thus we cannot simply take the $t\to0$ limit in~Eq.~\eqref{eq:(2.29)} because
of the presence of the identity operator. To avoid this, we define the
renormalized scalar density by the small flow-time limit of the chiral
transformation of the pseudo-scalar density~\eqref{eq:(2.15)} after the vacuum
expectation value subtracted. That is, we set
\begin{align}
   &\left\{\Bar{\psi}\{\{t^A,M\},t^B\}\psi\right\}_R(x)
\notag\\
   &\equiv\lim_{t\to0}\left\{1+
   \frac{\Bar{g}(1/\sqrt{8t})^2}{(4\pi)^2}C_2(R)
   \left[3\ln\pi+2
   +\ln(432)\right]\right\}
\notag\\
   &\qquad{}
   \times
   \left[
   \mathring{\Bar{\chi}}(t,x)\{\{t^A,\Bar{M}(1/\sqrt{8t})\},t^B\}
   \mathring{\chi}(t,x)
   -\left\langle
   \mathring{\Bar{\chi}}(t,x)\{\{t^A,\Bar{M}(1/\sqrt{8t})\},t^B\}
   \mathring{\chi}(t,x)
   \right\rangle
   \right].
\label{eq:(2.30)}
\end{align}
One might think that this subtraction, which makes the chiral condensation in
the vacuum, $\langle\{\Bar{\psi}\{\{t^A,M\},t^B\}\psi\}_R(x)\rangle$, always
vanishing is rather ad hoc. We should note, however, that the flavor singlet
part of the scalar density possesses the quantum number identical to the
identity operator and its $c$-number component (i.e., the mixing with the
identity operator) is ambiguous without supplementing a certain prescription
(except for the chiral limit~$M\to0$). It appears that the
definition~\eqref{eq:(2.30)} is the most natural one for massive fermions from
this perspective. For the chiral limit~$M\to0$, for which the mixing with the
identity operator in~Eq.~\eqref{eq:(2.29)} vanishes for dimensional reason,
we may adopt another definition of the scalar density (setting $M=mT$ with a
constant matrix~$T$),
\begin{align}
   &\lim_{m\to0}\left\{\Bar{\psi}\{\{t^A,T\},t^B\}\psi\right\}_R(x)
\notag\\
   &\equiv\lim_{t\to0}\lim_{m\to0}\left\{1+
   \frac{\Bar{g}(1/\sqrt{8t})^2}{(4\pi)^2}C_2(R)
   \left[3\ln\pi+2
   +\ln(432)\right]\right\}
\notag\\
   &\qquad\qquad\qquad{}
   \times
   \mathring{\Bar{\chi}}(t,x)\{\{t^A,\frac{\Bar{M}(1/\sqrt{8t})}{m}\},t^B\}
   \mathring{\chi}(t,x),
\label{eq:(2.31)}
\end{align}
which might be employed to compute the chiral condensation in the vacuum.

For the $N_f=2+1$ quantum chromodynamics (QCD), setting
\begin{equation}
   M=\begin{pmatrix}
   m_{ud}&0&0\\
   0&m_{ud}&0\\
   0&0&m_s\\
   \end{pmatrix},\qquad
   t^A=t^B=\frac{i}{2}\begin{pmatrix}
   0&1&0\\
   1&0&0\\
   0&0&0\\
   \end{pmatrix},
\label{eq:(2.32)}
\end{equation}
Eq.~\eqref{eq:(2.30)} gives (since $C_2(R)=4/3$ for the fundamental
representation of~$G=SU(3)$)
\begin{align}
   &\left\{\Bar{\psi}_u\psi_u\right\}_R(x)
   +\left\{\Bar{\psi}_d\psi_d\right\}_R(x)
\notag\\
   &=\lim_{t\to0}\left\{1+
   \frac{\Bar{g}(1/\sqrt{8t})^2}{(4\pi)^2}
   \frac{4}{3}\left[3\ln\pi+2+\ln(432)\right]\right\}
   \frac{\Bar{m}_{ud}(1/\sqrt{8t})}{m_{ud}}
\notag\\
   &\qquad\qquad\qquad\qquad{}\times
   \left[\mathring{\Bar{\chi}}_u(t,x)\mathring{\chi}_u(t,x)
   +\mathring{\Bar{\chi}}_d(t,x)\mathring{\chi}_d(t,x)-\text{VEV}\right].
\label{eq:(2.33)}
\end{align}
Similarly, by setting
\begin{equation}
   t^A=t^B=\frac{i}{2}\frac{1}{\sqrt{3}}\begin{pmatrix}
   1&0&0\\
   0&1&0\\
   0&0&-2\\
   \end{pmatrix},
\label{eq:(2.34)}
\end{equation}
and using Eq.~\eqref{eq:(2.33)}, we have
\begin{align}
   \left\{\Bar{\psi}_s\psi_s\right\}_R(x)
   &=\lim_{t\to0}\left\{1+
   \frac{\Bar{g}(1/\sqrt{8t})^2}{(4\pi)^2}
   \frac{4}{3}\left[3\ln\pi+2+\ln(432)\right]\right\}
   \frac{\Bar{m}_s(1/\sqrt{8t})}{m_s}
\notag\\
   &\qquad\qquad\qquad\qquad{}\times
   \left[\mathring{\Bar{\chi}}_s(t,x)\mathring{\chi}_s(t,x)-\text{VEV}\right].
\label{eq:(2.35)}
\end{align}

The representations~\eqref{eq:(2.33)} and~\eqref{eq:(2.35)} provide a possible
method to compute the chiral condensation at finite temperature, as
demonstrated recently by the WHOT QCD collaboration for the $N_f=2+1$
QCD~\cite{Taniguchi:2016}.

\section{Flavor singlet axial-vector current and the topological charge density}
\label{sec:3}
The flavor singlet axial-vector current~$j_{5\mu R}(x)$ requires a special
treatment because its renormalization is ambiguous due to the axial (or chiral)
anomaly. The chiral WT relation associated with the flavor-singlet chiral
symmetry is
\begin{align}
   &\left\langle
   \left[
   \partial_\mu j_{5\mu R}(x)
   -2\left\{\Bar{\psi}\gamma_5M\psi\right\}_R(x)+4N_fT(R)q_R(x)
   \right]
   \prod_i\psi(y_i)\prod_j\Bar{\psi}(z_j)\prod_kA_{\mu_k}^{a_k}(w_k)
   \right\rangle
\notag\\
   &=-\sum_l\delta(x-y_l)
   \left.\left\langle
   \prod_i\psi(y_i)\prod_j\Bar{\psi}(z_j)\prod_kA_{\mu_k}^{a_k}(w_k)
   \right\rangle\right|_{\psi(y_l)\to\gamma_5\psi(y_l)}
\notag\\
   &\qquad\qquad{}
   -\sum_l\delta(x-z_l)
   \left.\left\langle
   \prod_i\psi(y_i)\prod_j\Bar{\psi}(z_j)\prod_kA_{\mu_k}^{a_k}(w_k)
   \right\rangle\right|_{\Bar{\psi}(z_l)\to\Bar{\psi}(z_l)\gamma_5},
\label{eq:(3.1)}
\end{align}
where $q_R(x)$ stands for the topological charge density induced by the axial
anomaly.

\begin{figure}[ht]
\centering
\includegraphics[width=0.5\columnwidth]{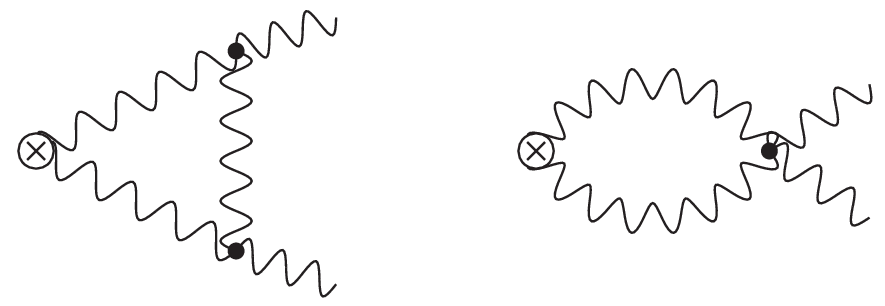}
\caption{One loop diagrams which contribute to the renormalization of the
topological charge density which is denoted by the blob.}
\label{fig:3}
\end{figure}

\begin{figure}[ht]
\centering
\includegraphics[width=0.2\columnwidth]{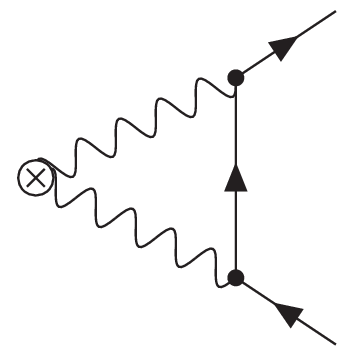}
\caption{One loop diagram which gives the operator mixing between the
topological charge density and the axial vector current.}
\label{fig:4}
\end{figure}

An analysis of one-loop diagrams, Figs.~\ref{fig:2}--\ref{fig:4}, by employing
the dimensional regularization shows that the following combinations
\begin{align}
   j_{5\mu R}(x)
   &=\left[1+\frac{g^2}{(4\pi)^2}C_2(R)(-4)+O(g^4)\right]
   \Bar{\psi}(x)\gamma_\mu\gamma_5\psi(x),
\label{eq:(3.2)}
\\
   \left\{\Bar{\psi}\gamma_5M\psi\right\}_R(x)
   &=\left[1+\frac{g^2}{(4\pi)^2}C_2(R)(-8)+O(g^4)\right]
   \Bar{\psi}(x)\gamma_5M_0\psi(x),
\label{eq:(3.3)}
\\
   q_R(x)&=\mu^{D-4}\left[1+O(g^4)\right]\frac{1}{64\pi^2}
   \epsilon_{\mu\nu\rho\sigma}F_{\mu\nu}^a(x)F_{\rho\sigma}^a(x)
\notag\\
   &\qquad{}
   +\left[\frac{g^4}{(4\pi)^4}C_2(R)
   \left(-3\frac{1}{\epsilon}+\text{finite}\right)
   +O(g^6)\right]\partial_\mu[\Bar{\psi}(x)\gamma_\mu\gamma_5\psi(x)],
\label{eq:(3.4)}
\end{align}
are finite and fulfill the WT relation~\eqref{eq:(3.1)}. For
Eq.~\eqref{eq:(3.1)} to be meaningful for any~$D$, we have required that $q_R$
in~Eq.~\eqref{eq:(3.4)} is of mass dimension~$D$ and supplemented the
factor~$\mu^{D-4}$ by using the renormalization scale.
Equation~\eqref{eq:(3.4)} shows that the flavor-singlet axial vector current
requires infinite renormalization from the two-loop order through the axial
anomaly~\cite{Adler:1969gk}, a property quite different from the flavor
non-singlet axial vector current.

As far as the finiteness and the WT identity~\eqref{eq:(3.1)} are concerned,
however, renormalized operators are to a large extent arbitrary. That is, the
redefinition of renormalized operators,
\begin{equation}
   j_{5\mu R}(x)\to[1-4N_fT(R)z]j_{5\mu R}(x),\qquad
   q_R(x)\to q_R(x)+z\partial_\mu j_{5\mu R}(x),
\label{eq:(3.5)}
\end{equation}
where $z$ is a finite constant of the order~$O(g^2)$, does not affect the
finiteness and the WT relation~\eqref{eq:(3.1)}. Thus, in what follows, we will
consider only the combinations which are independent of this ambiguity.
From~Eqs.~\eqref{eq:(3.2)}--\eqref{eq:(3.4)}, they are given by
\begin{align}
   &\partial_\mu j_{5\mu R}(x)+4N_fT(R)q_R(x)
\notag\\
   &=\left[1+\frac{g^2}{(4\pi)^2}C_2(R)(-4)
   +\frac{g^4}{(4\pi)^4}C_2(R)
   N_fT(R)\left(-12\frac{1}{\epsilon}+\text{finite}\right)+O(g^6)\right]
\notag\\
   &\qquad\qquad\times
   \partial_\mu[\Bar{\psi}(x)\gamma_\mu\gamma_5\psi(x)]
\notag\\
   &\qquad{}
   +N_fT(R)\mu^{D-4}\left[1+O(g^4)\right]\frac{1}{16\pi^2}
   \epsilon_{\mu\nu\rho\sigma}F_{\mu\nu}^a(x)F_{\rho\sigma}^a(x),
\label{eq:(3.6)}
\\
   &\left\{\Bar{\psi}\gamma_5M\psi\right\}_R(x)
   =\left[1+\frac{g^2}{(4\pi)^2}C_2(R)(-8)+O(g^4)\right]
   \Bar{\psi}(x)\gamma_5M_0\psi(x).
\label{eq:(3.7)}
\end{align}
We now express bare operators on the right-hand sides of these expressions by
the small flow-time expansion of composite operators of flowed fields.

Let us start with the topological charge
density~$\epsilon_{\mu\nu\rho\sigma}F_{\mu\nu}^a(x)F_{\rho\sigma}^a(x)$. The
one-loop computation of the small flow-time expansion is straightforward and
we find
\begin{align}
   &\epsilon_{\mu\nu\rho\sigma}G_{\mu\nu}^a(t,x)G_{\rho\sigma}^a(t,x)
\notag\\
   &\stackrel{t\to0}{\sim}
   \left[1+O(g_0^4)\right]
   \epsilon_{\mu\nu\rho\sigma}F_{\mu\nu}^a(x)F_{\rho\sigma}^a(x)
\notag\\
   &\qquad{}
   +\frac{g_0^4}{(4\pi)^2}C_2(R)(-12)(8\pi t)^\epsilon
   \left(\frac{1}{\epsilon}+\frac{3}{2}\right)
   \partial_\mu\left[\Bar{\psi}(x)\gamma_\mu\gamma_5\psi(x)\right]
   +O(t)
\notag\\
   &=\left[1+O(g^4)\right]
   \epsilon_{\mu\nu\rho\sigma}F_{\mu\nu}^a(x)F_{\rho\sigma}^a(x)
\notag\\
   &\qquad{}
   +\mu^{4-D}\frac{g^4}{(4\pi)^2}C_2(R)(-12)
   \left[\frac{1}{\epsilon}+\ln(8\pi\mu^2t)+\frac{3}{2}\right]
   \partial_\mu\left[\Bar{\psi}(x)\gamma_\mu\gamma_5\psi(x)\right]
   +O(t).
\label{eq:(3.8)}
\end{align}
The computation of this expansion in the pure Yang--Mills theory is given in
Appendix~B of~Ref.~\cite{Suzuki:2015bqa}. By comparing the right-hand side
of~Eq.~\eqref{eq:(3.8)} and~Eq.~\eqref{eq:(3.4)}, we find that this is a
renormalized finite quantity. This is consistent with the fact that the
composite operator of the flowed gauge field on the left-hand side must be a
renormalized quantity. Substituting this into~Eq.~\eqref{eq:(3.6)}, we have
\begin{align}
   &\partial_\mu j_{5\mu R}(x)+4N_fT(R)q_R(x)
\notag\\
   &=\left[1+\frac{g^2}{(4\pi)^2}C_2(R)(-4)+O(g^4)\right]
   \partial_\mu[\Bar{\psi}(x)\gamma_\mu\gamma_5\psi(x)]
\notag\\
   &\qquad{}
   +N_fT(R)\mu^{D-4}\left[1+O(g^4)\right]\frac{1}{16\pi^2}
   \epsilon_{\mu\nu\rho\sigma}G_{\mu\nu}^a(t,x)G_{\rho\sigma}^a(t,x)+O(t).
\label{eq:(3.9)}
\end{align}
For the axial-vector current~$\Bar{\psi}(x)\gamma_\mu\gamma_5\psi(x)$ on the
right-hand side, the computation is identical to the one for the flavor
non-singlet axial vector current in~Eq.~\eqref{eq:(2.14)} except for the
absence of the group generator~$t^A$. Thus,
\begin{align}
   &\partial_\mu j_{5\mu R}(x)+4N_fT(R)q_R(x)
\notag\\
   &=\left\{1+\frac{g^2}{(4\pi)^2}C_2(R)
   \left[-\frac{1}{2}+\ln(432)\right]+O(g^4)\right\}
   \partial_\mu[\mathring{\Bar{\chi}}(t,x)\gamma_\mu\gamma_5
   \mathring{\chi}(t,x)]
\notag\\
   &\qquad{}
   +N_fT(R)\left[1+O(g^4)\right]\frac{1}{16\pi^2}
   \epsilon_{\mu\nu\rho\sigma}G_{\mu\nu}^a(t,x)G_{\rho\sigma}^a(t,x)+O(t),
\label{eq:(3.10)}
\end{align}
where we have set $D\to4$.

The computation for the pseudo-scalar density on the right-hand side
of~Eq.~\eqref{eq:(3.7)} is again identical to the one for~Eq.~\eqref{eq:(2.15)}
except for the absence of the group generator~$t^A$. We have
\begin{align}
   &\left\{\Bar{\psi}\gamma_5M\psi\right\}_R(x)
\notag\\
   &=\left\{1+\frac{g^2}{(4\pi)^2}C_2(R)
   \left[3\ln(8\pi\mu^2t)+2+\ln(432)\right]
   +O(g^4)\right\}
   \mathring{\Bar{\chi}}(t,x)\gamma_5M\mathring{\chi}(t,x)+O(t).
\label{eq:(3.11)}
\end{align}

Finally, repeating the renormalization group argument as in previous sections,
we obtain the desired expressions,
\begin{align}
   &\partial_\mu j_{5\mu R}(x)+4N_fT(R)q_R(x)
\notag\\
   &=\lim_{t\to0}
   \Biggl(\left\{1+\frac{\Bar{g}(1/\sqrt{8t})^2}{(4\pi)^2}C_2(R)
   \left[-\frac{1}{2}+\ln(432)\right]\right\}
   \partial_\mu[\mathring{\Bar{\chi}}(t,x)\gamma_\mu\gamma_5
   \mathring{\chi}(t,x)]
\notag\\
   &\qquad\qquad{}
   +N_fT(R)\frac{1}{16\pi^2}
   \epsilon_{\mu\nu\rho\sigma}G_{\mu\nu}^a(t,x)G_{\rho\sigma}^a(t,x)
   \Biggr),
\label{eq:(3.12)}
\\
   &\left\{\Bar{\psi}\gamma_5M\psi\right\}_R(x)
\notag\\
   &=\lim_{t\to0}\left\{1+
   \frac{\Bar{g}(1/\sqrt{8t})^2}{(4\pi)^2}C_2(R)
   \left[3\ln\pi+2+\ln(432)\right]
   \right\}
   \mathring{\Bar{\chi}}(t,x)\gamma_5\Bar{M}(1/\sqrt{8t})\mathring{\chi}(t,x).
\label{eq:(3.13)}
\end{align}

Let us discuss implication of the representation~\eqref{eq:(3.12)}. The
topological charge~$Q$, whose normalization is consistent with the chiral WT
relation~\eqref{eq:(3.1)}, is defined by
\begin{equation}
   Q\equiv\int d^4x\,\left[
   q_R(x)+\frac{1}{4N_fT(R)}\partial_\mu j_{5\mu R}(x)\right]
   =\int d^4x\,q_R(x),
\label{eq:(3.14)}
\end{equation}
According to our representation~\eqref{eq:(3.12)}, this can be expressed as
\begin{equation}
   Q=\lim_{t\to0}\int d^4x\,
   \frac{1}{64\pi^2}
   \epsilon_{\mu\nu\rho\sigma}G_{\mu\nu}^a(t,x)G_{\rho\sigma}^a(t,x).
\label{eq:(3.15)}
\end{equation}
On the other hand, it follows from the flow equation~\eqref{eq:(1.6)}
that~\cite{Luscher:2011}
\begin{equation}
   \partial_t\left[\epsilon_{\mu\nu\rho\sigma}G_{\mu\nu}^a(t,x)G_{\rho\sigma}^a(t,x)
   \right]=\partial_\mu W_\mu(t,x),\qquad
   W_\mu(t,x)
   \equiv4\epsilon_{\mu\nu\rho\sigma}D_\lambda G_{\lambda\nu}^a(t,x)G_{\rho\sigma}^a(t,x).
\label{eq:(3.16)}
\end{equation}
Since $W_\mu(t,x)$ is a \emph{gauge-invariant} current, we see that the
spacetime integral in~Eq.~\eqref{eq:(3.15)} is independent of the flow-time~$t$
and, for any~$t>0$,\footnote{Note that in this paper we are assuming that the
cutoff is sent to infinity after the renormalization. With a finite cutoff such
as a finite lattice spacing, the story is different and for instance the
topological charge~$Q$ can depend on the flow-time~$t$.}
\begin{equation}
   Q=\int d^4x\,
   \frac{1}{64\pi^2}
   \epsilon_{\mu\nu\rho\sigma}G_{\mu\nu}^a(t,x)G_{\rho\sigma}^a(t,x).
\label{eq:(3.17)}
\end{equation}
This is precisely the definition of the topological charge
advocated in~Ref.~\cite{Luscher:2010iy} and this definition has been
extensively employed recently to compute the topological susceptibility in
quenched and full QCD. Our representation~\eqref{eq:(3.12)} confirms that the
topological charge~\eqref{eq:(3.17)} is in fact normalized in a consistent
manner with the chiral WT relation~\eqref{eq:(3.1)}; this statement can also be
found in~Ref.~\cite{Ce:2015qha}.

\section{Conclusion}
\label{sec:4}
In this paper, we constructed small flow-time representations of various
fermion bilinear operators of mass dimension~$3$, such as the axial-vector
and vector currents, the pseudo-scalar and scalar densities, flavor non-singlet
and flavor singlet ones. We believe that these representations, not only
provide universal expressions being independent of regularization, also
practically are useful in actual lattice Monte Carlo simulations. We treated
various operators case by case, but we found a very simple pattern in the
representations: For vector and axial-vector currents, we replace
$\psi(x)\to\mathring{\chi}(t,x)$,
$\Bar{\psi}(x)\to\mathring{\Bar{\chi}}(t,x)$, and multiply it by the factor
\begin{equation}
   1+\frac{\Bar{g}(1/\sqrt{8t})^2}{(4\pi)^2}C_2(R)
   \left[-\frac{1}{2}+\ln(432)\right],
\end{equation}
and consider the $t\to0$ limit. For the scalar and pseudo-scalar densities,
after $\psi(x)\to\mathring{\chi}(t,x)$,
$\Bar{\psi}(x)\to\mathring{\Bar{\chi}}(t,x)$, we multiply it by the
factor\footnote{For the $\overline{\text{MS}}$ scheme, the factor $\ln\pi$ is
replaced by $\gamma-2\ln2$.}
\begin{equation}
   1+\frac{\Bar{g}(1/\sqrt{8t})^2}{(4\pi)^2}C_2(R)
   \left[3\ln\pi+2+\ln(432)\right],
\end{equation}
and replace the mass matrix~$M$ by the running one~$\Bar{M}(1/\sqrt{8t})$, and
consider the $t\to0$ limit; for the flavor singlet scalar density, before the
$t\to0$ limit, we subtract the vacuum expectation value or take the
limit~$M\to0$. Although there should exist an underlying mechanism for the
above simple ``universal'' rule, at this moment we do not have a convincing
understanding. Presumably, the WT relations in terms of the flowed fermion
fields provide the clue.

\section*{Acknowledgments}
We would like to thank
Shinji Ejiri,
Kazayuki Kanaya,
Masakiyo Kitazawa,
Hiroki Makino,
Yusuke Taniguchi,
Takashi Umeda,
and especially 
Kazuo Fujikawa
for valuable discussions which motivated this work.
H.~S. would like to thank the Department of Physics at National
Taiwan University, where a part of this work was carried out, for their kind
hospitality.
The work of H.~S. is supported in part by JSPS Grants-in-Aid for Scientific
Research Grant Number~JP16H03982.

\end{document}